\documentstyle[prl,aps,twocolumn,EPSF]{revtex}
\newcommand{\be}{\begin{equation}}
\newcommand{\ee}{\end{equation}}
\newcommand{\bea}{\begin{eqnarray}}
\newcommand{\eea}{\end{eqnarray}}

\begin{document}
\title{The microscopic model for the magnetic subsystem in HoNi$_2$B$_2$C}
\author{V.~K.~Kalatsky$^a$ and V.~L.~Pokrovsky$^{a,b}$}

\address{
$^a$ Department of Physics, Texas A\& M University, 
College Station, Texas 77843-4242
}
\address{
$^b$ Landau Institute for Theoretical Physics, 
Kosygin str.2, Moscow 117940, Russia
}

\date{DRAFT (version 2, August 6 1997)}
\maketitle
\abstract{We demonstrate that the system of localized magnetic
moments in HoNi$_2$B$_2$C can be described by the 4-positional
clock model. This model, at a proper choice of the coupling constants, 
yields several metamagnetic phases in magnetic field at zero
temperature in full agreement with the experimental phase diagram.
The model incorporates couplings between not nearest neighbors
in the direction perpendicular to the ferromagnetic planes.
The same model leads to a c-modulated magnetic phase near the
Curie temperature. The theoretical value of the modulation 
wave-vector agrees surprisingly well with that observed by the
neutron diffraction experiment without new adjustable parameters.}

\vspace{-0.5cm}
In the works by Rathnayaka {\em et al.}~\cite{Naugle1} 
and by Canfield {\em et al.}~\cite{Canfield1} 
transport and magnetic measurements on HoNi$_2$B$_2$C for various 
magnetic fields and low temperatures have been reported. 
The magnetic phase diagram for HoNi$_2$B$_2$C with fields in the 
$a$-$b$ plane is of particular interest. 

In this compound easy magnetization axes are identified 
with crystallographic directions $\langle 110\rangle$ and 
$\langle1\overline{1}0\rangle$. The low temperature magnetization 
data show the existence of 4 metamagnetic phases. The low field 
phase has been identified by neutron diffraction 
experiments~\cite{neutron1,neutron3} and magnetic 
measurements~\cite{Canfield1} with the antiferromagnetic phase, 
which we denote symbolically $\uparrow\downarrow$. 
The phase boundaries 
and magnetization in other phases versus magnetic field 
found in the experiment~\cite{Canfield1} can be 
readily explained by assuming that the remaining three phases 
are as follows: phase 2 -- $\uparrow\uparrow\downarrow$, 
phase 3 -- $\uparrow\uparrow\rightarrow$, and the high-field 
phase 4 -- $\uparrow$. It means that $\frac23$ of the spins 
in the phases 2 and 3 are parallel to one of the easy axes 
whereas the remaining $\frac13$ is antiparallel and 
perpendicular, respectively, to the same axis.  Note that all metamagnetic 
phases are stoichiometric in the meaning that the concentrations 
of spins parallel, antiparallel, or perpendicular to the reference axis are 
rational numbers.
The phase diagram of HoNi$_2$B$_2$C at zero temperature is 
especially simple if the components of magnetic field $H_x$, 
$H_y$ are chosen as variables. 
The experimental phase diagram of HoNi$_2$B$_2$C is shown in  
Fig.~1 (in the original work~\cite{Canfield1} it has 
been presented in polar coordinates $h$, $\theta_h$). 
    
Siegrist  {\it et al.} \cite{siegrist} and Huang  {\it et al.}
\cite{neutron4} determined the structure of Lu and Ho 1:2:2:1
compounds as the body-centered tetragonal lattice with the space
group $I4/mmm$.
The  x-ray structure analysis and the neutron-scattering 
experiments 
performed by Goldman {\em et al.}~\cite{neutron1,neutron2} 
and Grigereit {\em et al.}~\cite{neutron3,neutron4,neutron5}.
showed that an incommensurate modulated magnetic structures
with the wave-vectors ${\bf K}_c=0.915{\bf c}^*$ and 
${\bf K}_a=0.585{\bf a}^*$ occur in the temperature range 4.7-6~K.
At temperatures below 4.7~K they vanish and an antiferromagnetic
reflections corresponding to alternating ferromagnetic $ab$-planes
of Ho$^{3+}$ localizad moments appear. Though the spacial arrangement
of the phases $\uparrow\uparrow\downarrow$ and $\uparrow\uparrow\rightarrow$
can not bee directly derived from the magnetization measurements,
it is unplausible that the ferromagnetic in-plane interaction changes
suddenly by switching on of the magnetic field. Therefore, we believe
that our symbols $\uparrow\uparrow\downarrow$ and $\uparrow\uparrow\rightarrow$
correspond to the real spatial sequences of in-plane magnetic moments.
\vspace{-0.5cm} 
\begin{figure}\protect
\centerline{\epsfxsize=3in\epsfysize=3in
\epsffile{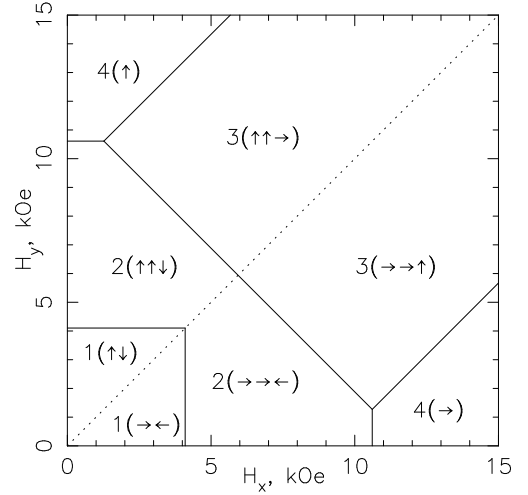}}
\caption{Magnetic phase diagram for HoNi$_2$B$_2$C. 
H$_x$ axis correspond to $\langle110\rangle$ direction.}
\label{fig:1}
\end{figure}  

In this article we present a simple microscopic model for magnetic subsystem
in the 1:2:2:1 compound which explains all experimental observations. 
We accept a model of strong anisotropy in which a single ion moment is directed
presumably along 4 easy directions ($\pm (1,1,0),\,\pm (1,{\overline 1},0)$
for the Ho and Dy compounds). Thus, the initially continuous moment $J$
is reduced to a discrete variable taking only 4 values. This is a kind of the 
so-called clock model with 4 positions of the "hand". 

The main argument in favor of the clock model is that 
the saturation magnetization in  the range of fields 
larger than $7-10T$ is directed not along the field, 
but along the closest to the field easy direction. 
It means that the applied field us still smaller than the 
anisotropy field $H_A$. The latter can be roughly estimated as $60T$. 
The corresponding anisotropy energy for Ho$^{3+}$ magnetic
moment $\sim 10\mu_B$ is about $40meV\approx 400K$. 
Nevertheless, a single ion with integer $J$ has no average moment 
in the ground state in the absence of external magnetic field.
It happens since even a small tunneling amplitude $w$ between 
adjacent positions of the "hand" (moment) leads to 
the ground-state in which all 4 positions have equal probabilities. 
The ground-state is separated by a finite energy gap equal to $2|w|$
from the first excited state. A detailed analysis of the single-ion properties
including the difference between integer and half-integer $J$ will be 
published separately. Here we focus on the description of collective effects.

For this purpose we introduce 
an angular variable $\theta_{\bf r}$
at any lattice site $\bf r$ taking independently 4 values 
$0,\,\pi/2,\,\pi,\,3\pi/2$.
Neglecting the tunneling, the most general Hamiltonian 
compatible with the tetragonal symmetry is:
\bea
H&=&{1\over 2}\sum_{\bf r, r^{\prime}}\left[ 
K({\bf r-r}^{\prime})\cos (\theta_{\bf r}
-\theta_{\bf r}^{\prime})+L({\bf r-r}^{\prime})\cos 2(\theta_{\bf r}
-\theta_{\bf r}^{\prime})\right]\nonumber\\
&-&h_x\sum_{\bf r}\cos\theta_{\bf r}-h_y\sum_{\bf r}\sin\theta_{\bf r}
\label{coll}
\eea
where $K({\bf r})$ and $L({\bf r})$ are coupling constants 
and $h_{x,y}$ are components of the magnetic field. 
We employ the reference frame in which axes coincide with the
easy axis directions. 

The higher harmonic terms are generated by the exchange interaction.
Indeed, the operator of two particle permutation specific for the exchange 
is a linear function of the scalar product of particle spins for spin 1/2.
For higher spins $J$ this operator contains higher powers of the same 
scalar product up to $2J$. The dipolar interaction in principle is 
not small for the Ho compound since the magnetic moment of 
Ho$^{3+}$ is 
large (about $10 \mu_B$). However, it is effectively reduced in
metamagnetic systems. The dipolar interaction 
between ferromagnetically aligned planes  
is proportional to a small factor $\exp{(-2\pi c/a)}$, where $c$
is interplane and $a$ is in-plane lattice constants.
Returning to the exchange interaction, we find that for the
4-positional spins in plane only the invariants 
${\bf S}_1{\bf S}_2=\cos(\varphi_1-\varphi_2)$ and
$({\bf S}_1{\bf S}_2)^2=\left(\cos(\varphi_1-\varphi_2)\right)^2$
are independent, all the rest $2J-3$ invariants are equal to one of
these two.

The tetragonal symmetry of the single ion ground-state is violated spontaneously 
if $\max{K({\bf r})}>|w|$. This intuitive idea is supported by a variational
calculation. As a result a Neel state arises with non-zero
magnetic moments on Ho$^{3+}$ ions.

Let us restrict the set of coupling constants to a few
independent values.  We assume that the in-plane interaction is
characterized by one nearest-neighbor negative constant $K$ with
all other in-plane $K_{\bf r}$ and all in-plane $L_{\bf r}$ equal to zero.  
The in-plane interaction is assumed to be dominant to provide the
in-plane ferromagnetic order.  The interplane interaction is
characterized by several constants $K_n, L_n$. We shall see that 
interaction with several neighbors is essential, not only the nearest 
neighbor interaction.

All spins in each plane are parallel. Thus, the
ground state is determined by minimization of a spin-chain
Hamiltonian:
\begin{equation}
H=\!\!\!\sum_{i,=-\infty, n=1}^{\infty}\!\!\! 
\left[ K_n\,\cos{(\varphi_i-\varphi_{i+n})}\!+\!
L_n\,\cos{2(\varphi_i-\varphi_{i+n})}\right]
\label{chain}
\end{equation}
It should be noted that in the
absence of an applied magnetic field it is known that the N\'{e}el
antiferromagnetic state consists of alternating ferromagnetic $a$-$b$
planes.  This requirement is satisfied, if $K_1 > 0$.
A natural desire to simplify the model leaving one or two independent
coupling constants unfortunately cannot be fulfilled. For example, 
if one leaves 
non-zero $K_1$ and  $K_2$ and puts $L_1$, $ L_2$ and all the rest
$K_n$, $L_n\,\,(n\geq 3)$ to be zero, two
kind of the phase diagrams occur.  The first diagram, Fig.~2,
corresponds to $0 < K_2 < K_1/2$ (the latter inequality is necessary
to have the antiferromagnetic state in zero field).  It contains 6
phases with net distribution of the in-plane moments symbolically
depicted.  Due to the symmetry only the sector $0 < h_y < h_x$ must
be considered.  Fig.~3 corresponds to $K_2 < 0$. It is simpler
and contains only 3 phases.  Neither of the phase diagrams fits
the experiment which clearly displays 4 phases with different
properties as shown in Fig.~1.  This shows that at least 
the coefficients 
$L_1$ and $L_2$ must be incorporated to describe the experimental
situation in HoNi$_2$B$_2$C.  Moreover, we shall show later that the coefficient 
$L_3$ is not zero. Thus, we restrict our model to six non-zero
coupling constants $K_n,\,L_n,\,\,\,\,n=1,\,2,\,3$. This is a generalization
of the so-called anisotropic next-nearest neighbor Ising (ANNNI) 
model \cite{bak}, \cite{fischer-selke}. Our
model differs from the standard ANNNI one by two features: the 
third-neighbor interaction and the 4 positions of the hand instead of 
two in the Ising model.
\vspace{-0.8cm}
\begin{figure}\protect
\centerline{\epsfxsize=3in\epsfysize=3in\epsffile{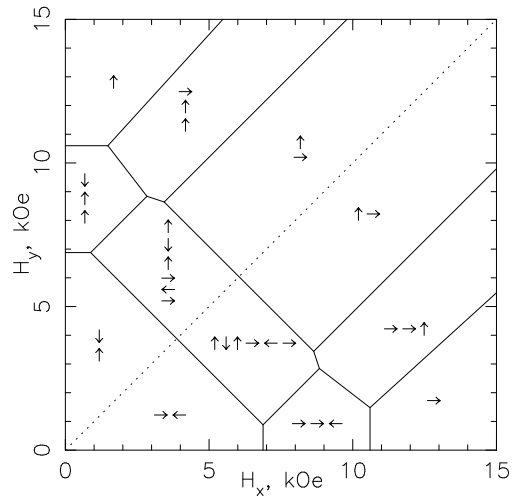}}
\caption{Magnetic phase diagram. All the parameters are the same as
in Fig. 1 except $L_2=0.05$, $K_3=L_3=0$.}
\label{fig:3}
\end{figure}
\vspace{-1cm}
\begin{figure}
\centerline{\epsfxsize=3in\epsfysize=3in\epsffile{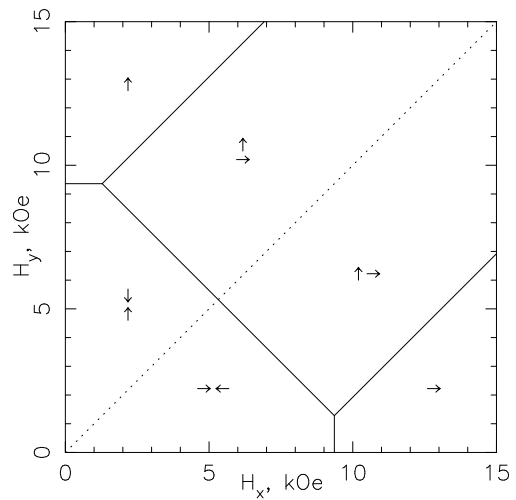}}
\caption{Magnetic phase diagram. All the parameters are the same as
in Fig. 1 except $K_2=-0.62$, $K_3=L_3=0$.}
\label{fig:4}
\end{figure}

In order to understand why the second and third neighbor interaction must
be incorporated, one should compare energies of the simplest periodic
sequences in the chain. The phases we anticipate
to be realized as the ground states at different values of the
field $\bf h$ are: $\uparrow\downarrow$ (AF, period 2 (layers)), 
$\uparrow$ (F, period 1), 
$\uparrow\uparrow\downarrow$ and $\uparrow\uparrow\rightarrow$ (period 3).
Others, having rather close energies, are $\uparrow\rightarrow$ 
(period 2); $\uparrow\uparrow\uparrow\downarrow$, 
$\uparrow\uparrow\uparrow\rightarrow$, 
$\uparrow\uparrow\downarrow\rightarrow$, and 
$\uparrow\downarrow\uparrow\rightarrow$ (period 4); 
$\uparrow\uparrow\downarrow\uparrow\downarrow$,
$\uparrow\uparrow\uparrow\uparrow\downarrow$ and
$\uparrow\uparrow\rightarrow\uparrow\rightarrow$ (period 5);
$\uparrow\uparrow\uparrow\uparrow\uparrow\downarrow$ and
$\uparrow\downarrow\uparrow\rightarrow\leftarrow\rightarrow$
(period 6). We have found by numerical sorting that other phases
have larger energies and can be omitted. With these 14 phases
participating in the competition, a number of inequalities
must be satisfied to ensure the existence of the experimentally
observed phase diagram. Namely, on the phase boundaries
the energies of the phases other than those being in equilibrium
must be larger. For the reader's convenience the
energies of the competing 14 phases are given in Table I.
All are linear functions of the magnetic field. Therefore
only their values at the corners of the phase diagram should 
be compared.

The general investigation of the phase diagram in the 8-dimensional
space of $K_n, L_n$ and $h_x, h_y$ is too cumbersome. Instead we
assume that the phase diagram has 4 phase boundaries, separating
the experimentally established 4 phases, and find what are the
constraints imposed by the experiment onto the model. The 
four phase boundaries found in the experiment are: 
$$
AF\,\leftrightarrow\,\uparrow\uparrow\downarrow
\,\leftrightarrow\,\uparrow\uparrow\rightarrow
\,\leftrightarrow\,F
\,\leftrightarrow\,\uparrow\uparrow\downarrow.
$$
According to the Table I, these lines are described by the 
following equations, in the same respective order as above:
\begin{eqnarray}
h_x\,&=&\,2(K_1\,-\,2K_2\,+\,3K_3)\equiv H_{c10};
\label{pt1}\\
h_x+h_y\,&=&\,2(K_1+K_2)\,-\,4(L_1+L_2)\equiv\sqrt{2}H_{c20};
\label{pt2}\\
h_x-h_y\,&=&\,2(K_1+K_2)\,+\,4(L_1+L_2)\equiv\sqrt{2}H_{c30};
\label{pt3}\\
h_x\,&=&\,2(K_1+K_2).
\label{pt4}
\end{eqnarray}
The latter three lines intersect in the triple point
$h_x^t=2(K_1+K_2)$, $h_y^t=-4(L_1+L_2)$.
Note that Eqs. (\ref{pt1}-\ref{pt3}) are equivalent to empirical 
equations for the transition lines found 
in~\cite{Canfield1}. 
Thus, the theory suggests a natural explanation of all
functional dependencies, $H_{c1}(\theta ),\, H_{c2}(\theta )$, and 
$H_{c3}(\theta )$, found in the experiment. The phase diagram in 
the plane $h_x, h_y$ has an extremely simple shape (see Fig.~1). 

It is worthwhile to note that all above discussed functional 
dependencies were derived from purely geometrical
considerations.
However, the very existence
of the phase diagram with the four phases observed in the experiment
is highly nontrivial and imposes strong constraints on the 
coupling constants. These constraints are expressed as a long
series of inequalities. We present here the four most important
of them with necessary comments on their meaning:
\begin{itemize}
\item{$K_1\,-\,2K_2\,+\,3K_3>0$. AF phase has minimal energy at
$h_x<H_{c10}$.}
\item{$K_1-2K_2+3K_3+2(L_1+2L_2)+6L_3<0$. The AF and 
$\uparrow\uparrow\downarrow$ phases have lower energy than the phase
$\uparrow\downarrow\uparrow\rightarrow\leftarrow\rightarrow$ on
the phase boundary $h_x=H_{c10}$.}
\item{$K_1-2K_2+3K_3+2(L_1-2L_2)+6L_3<0$. The phases 
$\uparrow\uparrow\downarrow$ and $\uparrow\uparrow\rightarrow$ have
lower energy than the phase $\uparrow\rightarrow$ on the
phase boundary $h_x+h_y=\sqrt{2}H_{c20}$.}
\item{$2K_1-4K_2+9K_3+4(L_1+L_2)+6L_3<0$. The phases
$\uparrow\uparrow\downarrow$ and $\uparrow\uparrow\rightarrow$ have
lower energy than the phase $\uparrow\downarrow\uparrow\rightarrow$ on the
phase boundary $h_x+h_y=\sqrt{2}H_{c20}$.}
\end{itemize}
One can deduce from the second and third inequalities that: 
$$ K_1\,-\,2K_2\,+\,3K_3\,+\,2(L_1+L_2)\,+\,6L_3\,<\,0.$$
Employing equations (\ref{pt1}-\ref{pt3}) and the third
inequality we obtain the following inequality:
\begin{equation}
L_3 < -\left( \frac{H_{c01}}{12}\,-
\,\frac{H_{c20}-H_{c30}}{12\sqrt{2}}\right).
\label{L3-limit}
\end{equation}
With the experimental values
$H_{c10}=4.1$ kG, $H_{c20}=8.4$ kG and $H_{c30}=6.6$ kG 
in inequality (\ref{L3-limit}), we find
the upper boundary for $L_3$: $L_3< - 0.24$ kG. Thus, we see that
the coupling constant $L_3$ cannot be zero.
Thus, the experimental data imply that the interaction between magnetic
planes separated by 3 half-periods $3c/2$ is essential. Taking this
interaction into account, we otherwise follow the principle of minimal
interaction. It means that we put as many as possible coupling constants
to be zero. In particular, we put $K_3=L_2=0$.

In the framework of our rough theory the magnetization in 
each phase does not depend on the magnetic field. 
It is equal to zero in the AF ($\uparrow\downarrow$) phase. 
In the phase $\uparrow\uparrow\downarrow$ 
it is directed along an easy axis 
closest to the direction of the magnetic field, 
and its absolute value is equal to 1/3 of the easy-axis 
saturation value. 
In the phase $\uparrow\uparrow\rightarrow$ the magnetization is
tilted by an angle $\arctan (1/2) = 26.6^{\circ}$ 
to the easy axis closest to the magnetic field, and 
its absolute value is equal to $\sqrt{5}/3=0.745$ of the
easy-axis saturation value. In the ferro-phase $\uparrow$ 
it is equal to 1 per site.
In the experiment~\cite{Canfield1} the projection of magnetization onto the 
field direction was measured. 
According to the theory it is $(1/3)\cos\theta_h$ 
for the phase $\uparrow\uparrow\downarrow$
(phase 2), $0.745\cos (\theta -26.6^{\circ})$ in the phase 
$\uparrow\uparrow\rightarrow$ (phase 3)
and $\cos\theta$ in the ferro-phase. 
While theoretical values of the magnetization
in the phase 2 and the ferro-phase are in a good agreement 
with the experimental data, there is a discrepancy between 
the theoretical and experimental magnetization
of the phase 3 (see Fig. 5(c), in~\cite{Canfield1}). 
In particular, in the experiment there is no maximum
of $M_{s2}(\theta)$ at $\theta =26.6^{\circ}$ as the theory predicts. 
Instead the saturation magnetization decreases monotonically with 
the angle in the interval $15^{\circ}<\theta < 45^{\circ}$. 
The reason can be that the determination of the $M_{s2}$ at small 
angle is very unreliable since the plateau is not clearly pronounced.
On the other hand the values of magnetization at orientations 
closer to the easy axis where the plateau is well pronounced 
are in a good agreement with the theory. 
Finally, the relative difference of the magnetization at a maximum 
($\theta=26.6^{\circ}$) and at 
$\theta=45^{\circ}$ is only 5\% which may be beyond of the 
precision of the model without the tunneling taken into 
account ($w=0$). 

From equations
(\ref{pt1}-\ref{pt3}) one can find: 
\be
K_1\,=\,4.22\,\mbox{kG};\,\,\,\,\,\,K_2\,=\,1.08\,\mbox{kG};\label{K} 
\ee
\be
L_1\,=\,-0.32\,\mbox{kG};\,\,\,\,\,\,L_3\,=\,-0.46\,\mbox{kG}. \label{L} 
\ee
Thus, we demonstrated that the low-temperature 
magnetization data and corresponding
phase diagram can be naturally described in the frameworks of 4-position
clock model withthe values of constants given by Eqns. (\ref{K}, \ref{L}).

Now we consider a vicinity of the Curie temperature. We will show that the
modulation along the c-direction naturally appears in 
the framework of the same model.
The order parameter (magnetization in a plane) is small 
near this temperature allowing one to neglect the terms with the 
$\cos{2(\theta_n-\theta_n^{\prime})}$ 
in the Hamiltonian~(\ref{chain}) (they are proportional to the fourth power
of the order parameter $\bf s$).
The chain interaction Hamiltonian becomes:
\begin{equation}
H=\sum_{i=-\infty}^{\infty}\sum_{n=1}^3K_n {\bf s}_i\cdot{\bf s}_{i+n}.
\label{3neighbor}
\end{equation}
The quadratic Hamiltonian (\ref{3neighbor}) can be represented 
in terms of Fourier-components 
${\bf s}_q=N^{-1/2}\sum_{n=1}^Ne^{iqn}{\bf s}_n$:
\begin{equation}
H\,=\,\sum_qK_q{\bf s}_q{\bf s}_{-q}
\label{fourier}
\end{equation}
with $K_q=K_1\cos q + K_2\cos 2q$. The value $K_q$ has the absolute minimum at 
$q=\arccos (-K_1/4K_2)$, if $|K_1|<4|K_2|$. For our data $K_1/(4K_2)=0.977$
and $q=167^{\circ}=0.93c^{\ast}$. 
Comparing the theoretical value to the experimental
value $q=0.915c^{\ast}$, we find the agreement to be surprisingly good, may be
too good. 
We can introduce the constant $K_3$ to compensate a small discrepancy.
The value $K_3$ established in this way is $-0.023$. Though this value is not
reliable, it shows that our minimal value was close to reality.
No modulated magnetic phase has been found 
for the Dy compound \cite{Dy-neutron}. 
From our point of view it means that the ratio $K_1/4K_2$ is 
larger than 1 in this compound.

An important remark
is in order: several phases which do not occur in the phase diagram
have energies very close to the ground state energy. 
This means that a small perturbation (stress) can change 
the phase diagram drastically.

The next step toward a more realistic theory would be to incorporate 
the non-zero tunneling amplitude $w$. 
The CEF spectrum numerical calculations \cite{Harmon1} for this amplitude give the 
magnitude $w\approx 3$ kG, which is not small, 
especially in comparison to $L_1$ and $L_2$. 
The incorporation of the tunneling amplitude, 
probably weakens the strong limitations imposed by the inequalities. 
We have performed a variational calculation of the 
ground-state for the extended model including
$w$ in the Hartree approximation. Our calculation shows that for zero $L_i$
the energy has the property of separability: 
it is a sum of identical functions of $h_x$ and $h_y$. 
Therefore, the angular dependencies of $H_{c2}$ and $H_{c3}$
remains the same at least for small $L_i$ independently of the value of $w$.
The variational calculation will be published elsewhere. 

Another important and not resolved yet question is the origin and 
the behavior of the $a$-modulation with the wave-vector $0.585 a^{\ast}$. 
It appears not only in the Ho compound, but also in Er, Tm, and, 
Tb~\cite{Tb-neutron}. 
Its wave-vector is very conservative. 
Therefore, it is tempting to ascribe it to
a spin-density wave in the conductivity electrons. 
This idea is supported by an observation 
of good nesting on the numerically calculated Fermi-surface~\cite{Harmon2}. 
However, such a treatment does not agree with the fact that
in the Ho compound the $a$ and $c$-modulations appear and disappear
in the same temperature interval. Rathnayaka {\em et al.}~\cite{Naugle1} have
found an additional phase transition in the same temperature interval.
It can be considered as an implicit indication on the independence of these
order parameters. From the theoretical point of view there is no reason for
them to appear in the same point. 
However, direct neutron diffraction measurements
do not distinguish the temperture where these modulations appear.

\begin{center}
{ Table I. Competing phases and their energies.}

\begin{tabular}{||l|l||}
\hline
Phase & Energy of the phase\\
\hline
\hline
AF ($\uparrow\downarrow$) & $-K_1+K_2-K_3+L_1+L_2+L_3$\\
\hline
F ($\uparrow$) & $K_1+K_2+K_3+L_1+L_2+L_3-h_x$\\
\hline
$\uparrow\rightarrow$ & $K_2-L_1+L_2-L_3-{{h_x+h_y}\over 2}$\\
\hline
$\uparrow\uparrow\downarrow$ & $-{K_1\over 3}-{K_2\over3}+K_3
+L_1+L_2+L_3-{h_x\over 3}$\\
\hline
$\uparrow\uparrow\rightarrow$ & ${K_1\over 3}+{K_2\over 3}+K_3
-{L_1\over 3}-{L_2\over 3}+L_3-{{2h_x+h_y}\over 3}$\\
\hline
$\uparrow\uparrow\uparrow\downarrow$ &
$L_1+L_2+L_3-{h_x\over 2}$\\
\hline
$\uparrow\uparrow\uparrow\rightarrow$ &
$\frac{K_1}{2}+\frac{K_2}{2}+\frac{K_3}{2}-\frac{3h_x+h_y}{4}$ \\
\hline
$\uparrow\uparrow\downarrow\rightarrow$ &
$-\frac{K_2}{2}-\frac{h_x+h_y}{4}$\\
\hline
$\uparrow\downarrow\uparrow\rightarrow$ & 
$-{K_1\over 2}+{K_2\over 2}-{K_3\over 2}-{{h_x+h_y}\over4}$\\
\hline
$\uparrow\uparrow\uparrow\uparrow\downarrow$ &
${K_1\over 5}+{K_2\over 5}+{K_3\over 5}+L_1+L_2+L_3-{{3h_x}\over 5}$\\
\hline
$\uparrow\uparrow\downarrow\uparrow\downarrow$ &
$-{{3K_1}\over 5}+{K_2\over 5}+{K_3\over 5}+L_1+L_2+L_3-{h_x\over 5}$\\
\hline
$\uparrow\uparrow\rightarrow\uparrow\rightarrow$ &
${K_1\over 5}+{{3K_2}\over 5}+{{3K_3}\over 5}-{{3L_1}\over 5}+
{L_2\over 5}+{L_3\over 5}-{{3h_x+2h_y}\over 5}$\\
\hline
$\uparrow\uparrow\uparrow\uparrow\uparrow\downarrow$ &
${1\over 3}(K_1+K_2+K_3)+L_1+L_2+L_3-{{2h_x}\over 3}$\\
\hline
$\uparrow\downarrow\uparrow\rightarrow\leftarrow\rightarrow$ &
$-{{2K_1}\over 3}+{K_2\over 3}+{L_1\over 3}-{L_2\over 3}
-L_3-{{h_x+h_y}\over 6}$\\
\hline
\hline
\end{tabular}
\end{center}


\vspace{-1cm}
\begin{thebibliography}{99}
\bibitem{Naugle1} K.~D.~D.~Rathnayaka, D.~G.~Naugle, 
 B.~K.~Cho, and P.~C.~Canfield,
{\em Phys. Rev. B {\bf 53}, 5688 (1996)}.
\bibitem{Canfield1} P.~C.~Canfield, S.~L.~Bud'ko, B.~K.~Cho, 
A.~Lacerda, D.~Farrell, E.~Johnston-Halperin, 
V.~A.~Kalatsky, and V.~L.~Pokrovsky. 
{\em Phys. Rev. B {\bf 55}, 970 (1997)}.          
\bibitem{neutron1} A.~I.~Goldman, C.~Stassis, P.~C.~Canfield, 
J.~Zarestky, P.~Dervenagas, B.~K.~Cho, D.~C.~Johnston, and 
B.~Sternlieb, 
{\em Phys. Rev. B {\bf 50}, 9668 (1994)}.
\bibitem{neutron2} C.~Detlefs, A.~I.~Goldman, C.~Stassis, 
P.~C.~Canfield, and B.~K.~Cho, 
{\em Phys. Rev. B {\bf 53}, 3487 (1996)}.
\bibitem{neutron3} T.~E.~Grigereit, J.~W.~Lynn, Q.~Huang, 
A.~Santoro, R.~J.~Cava, J.~J.~Krajevski, and W.~E.~Peck, Jr., 
{\em Phys. Rev. Lett. {\bf 73}, 2756 (1994)}. 
\bibitem{neutron4} Q.~Huang, A.~Santoro, T.~E.~Grigereit, 
J.~W.~Lynn, R.~J.~Cava, J.~J.~Krajevski, and W.~E.~Peck, Jr., 
{\em Phys. Rev. B {\bf 51}, 3701 (1995)}.
\bibitem{neutron5} J.~W.~Lynn, Q.~Huang, A.~Santoro, R.~J.~Cava, 
J.~J.~Krajewski, and W.~E.~Peck, Jr.,
{\em Phys. Rev. B {\bf 53}, 802 (1996)}.
\bibitem{Naugle2} D.~C.~Naugle, K.~D,~D.~Rathnayaka, 
A.~K.~Bhatnagar, A.~C.~Du~Mar, A.~Parasiris, J.~M.~Bell, 
P.~C.~Canfield, and B.~K.~Cho, 
{\em Czech. J. Phys. {\bf 46} S6, 3263 (1996)}.
\bibitem{siegrist} T.~Siegrist, H.~W.~Zandbergen, R.~J.~Cava, 
J.~J.~Krajewski, and
W.~F.~Peck, Jr., {\em Nature {\bf 367}, 254 (1994)}.
\bibitem{*} This fact follows from the measurements of saturated magnetization
which do not display any tendency to the isotropization at field $H\sim 20$ T.
\bibitem{bak} P.~Bak, J.~von~Boehm, {\em Phys. Rev. B {\bf 21}, 5297 (1980)}.
\bibitem{fischer-selke} M.~E.~Fisher, W.~Selke, {\em Phys. Rev. Lett. 
{\bf 44}, 1502 (1980)}.
\bibitem{Dy-neutron} P.~Dervenagas, J.~Zarestky, C.~Stassis, A.~I.~Goldman,
P.~C.~Canfield, and B.~K.~Cho {\em Physica B {\bf 212}, 1 (1995)}.
\bibitem{Harmon1} B.~K.~Cho, B.~N.~Harmon, D.~C.~Johnston, 
and P.~C.~Canfield, {\em Phys. Rev. B {\bf 53}, 2217 (1996)}.
\bibitem{Tb-neutron} P.~Dervenagas, J.~Zarestky, C.~Stassis, A.~I.~Goldman, 
P.~C.~Canfield, and B.~K.~Cho {\em Phys. Rev. B {\bf 53}, 8506 (1996)}. 
\bibitem{Harmon2} J.~Y.~Rhee, X.~Wang, and B.~N.~Harmon 
{\em Phys. Rev. B {\bf 51}, 15585 (1995)}.
\end{thebibliography}
\end{document}